\documentclass[conference]{IEEEtran}

\usepackage[T1]{fontenc} 
\ifCLASSINFOpdf
\else
\fi
\usepackage{bm,color,soul}
\usepackage{amsmath, amsthm, amssymb}
\theoremstyle{plain}

\usepackage[pdftex]{graphics}
\theoremstyle{definition}

\usepackage{xspace}
\usepackage{amssymb}
\usepackage{graphicx}
\usepackage{amsmath}
\usepackage{mathtools}
\usepackage{graphicx,color} 
\usepackage{amsmath, amsfonts, amssymb, amsbsy,nccmath} 
\usepackage{algorithm} 
\usepackage{enumerate} 
\usepackage{algorithmic} 
\usepackage{lipsum} 
\usepackage[sort,compress]{cite} 
\usepackage{epsfig} 
\usepackage{epstopdf}
\usepackage{mathtools}
\usepackage[cmintegrals]{newtxmath}
\usepackage{color,soul}
\usepackage{stfloats}  
\usepackage{tabularx}
\usepackage{lipsum}

\usepackage{longtable}
\newcommand{\bmA}{\mathbf A}
\newcommand{\bma}{\mathbf a}
\newcommand{\bmI}{\mathbf I}
\newcommand{\bmT}{\mathbf T}

\newcommand{\bmB}{\mathbf B}
\newcommand{\bmC}{\mathbf C}
\newcommand{\bmD}{\mathbf D}
\newcommand{\bmE}{\mathbf E}

\newcommand{\bmG}{\mathbf G}
\newcommand{\bmH}{\mathbf H}
\newcommand{\bmF}{\mathbf F}
\newcommand{\bmS}{\mathbf S}
\newcommand{\bms}{\mathbf s}
\newcommand{\bmV}{\mathbf V}
\newcommand{\bmU}{\mathbf U}

\newcommand{\bmy}{\mathbf y}
\newcommand{\bmZ}{\mathbf Z}

\newcommand{\bmR}{\mathbf R}
\newcommand{\bmX}{\mathbf X}
\newcommand{\bmW}{\mathbf W}

\newcommand{\bmq}{\mathbf q}

\newcommand{\bmn}{\mathbf n}

\begin{document}
\title{Intelligent Reflecting Surfaces Assisted Millimeter Wave MIMO Full Duplex Systems}
\author{\IEEEauthorblockN{Chandan~Kumar~Sheemar\IEEEauthorrefmark{1}, Stefano Tomasin \IEEEauthorrefmark{2}, Dirk Slock\IEEEauthorrefmark{3}, and Symeon Chatzinotas\IEEEauthorrefmark{1}}
\IEEEauthorblockA{
\IEEEauthorrefmark{1}SnT, University of Luxembourg, email: \{chandankumar.sheemar, symeon.chatzinotas\}@uni.lu\\
\IEEEauthorrefmark{2}University of Padua, Italy, email: \{stefano.tomasin\}@unipd.it\\
\IEEEauthorrefmark{3}EURECOM, Sophia Antipolis, France, email: \{slock\}@eurecom.fr
}}
\maketitle
\begin{abstract}
In this paper, we propose to remove the analog stage of hybrid beamforming (HYBF) in the millimeter wave (mmWave) full-duplex (FD) systems. Such a solution is highly desirable as the analog stage suffers from high insertion loss and high power consumption. Consequently, the mmWave FD nodes can operate with a fewer number of antennas, instead of relying on a massive number of antennas, and to tackle the propagation challenges of the mmWave band we propose to use near-field intelligent reflecting surfaces (NF-IRSs). The objective of the NF-IRSs is to simultaneously and smartly control the uplink (UL) and downlink (DL) channels while assisting in shaping the SI channel: this to obtain very strong passive SI cancellation. A novel joint active and passive beamforming design for the weighted sum-rate (WSR) maximization for the NF-IRSs-assisted mmWave point-to-point FD system is presented. Results show that the proposed solution  fully reaps the benefits of the IRSs, only when they operate in the NF, which leads to considerably higher gains compared to the conventional massive MIMO (mMIMO) mmWave FD and half duplex (HD) systems.
\end{abstract}


\IEEEpeerreviewmaketitle

\section{Introduction} 
\IEEEPARstart{T}{o} achieve ubiquitous connectivity for continuously evolving wireless networks, intelligent reflecting surfaces (IRSs) are considered one of the most prominent hardware technology for beyond fifth generation (B5G) and sixth generation (6G) networks \cite{hu2018beyond}, with the potential to create smart, reconfigurable, and highly energy-efficient wireless systems. On the other hand, given the limited wireless spectrum, there has also been a growing interest in millimeter wave (mmWave) full-duplex (FD) technology, which offers simultaneous transmission and reception in the same frequency band, thus doubling the spectral efficiency with respect to  half duplex (HD) systems \cite{sheemar2022practical,rosson2019towards,sheemar2021massive}.
FD systems suffer from self-interference (SI), which needs to be cancelled to allow the correct reception of the useful signal. 

To perform FD operation in the millimeter wave (mmWave) band, a massive number of transmit and receive antennas are necessary to overcome the propagation challenges. Hybrid beamforming (HYBF) is considered a cost-efficient solution for such systems, to also partially perform SI cancellation (SIC) \cite{sheemar2021hybrid}. The mmWave FD systems aided by IRSs can enable highly spectrally efficient, energy efficient, smart, reconfigurable, and sustainable networks \cite{pan2020full}. Recent studies on FD systems, with IRSs, limited only to sub-$6$ GHz, are available in \cite{abdullah2020optimization,peng2021multiuser,sharma2020intelligent,cai2021intelligent}.
In \cite{abdullah2020optimization}, a novel beamforming design is proposed for  decode-and-forward FD relays assisted with IRS to maximize the minimum achievable rate. In \cite{peng2021multiuser}, the authors proposed a joint active and passive beamforming design to cover  dead zones in a multi-user multiple-input single-output (MISO) FD network, while suppressing the single-antenna FD user-side self and co-channel interference. In\cite{cai2021intelligent}, the authors presented a novel beamforming design for a MISO FD system with two users (one in uplink (UL) and one in downlink (DL)) to minimize the power consumption of the FD access point while the UL user is subject to the minimum rate constraint.

\begin{figure}
    \centering
    \includegraphics[width=6cm,height=3.5cm]{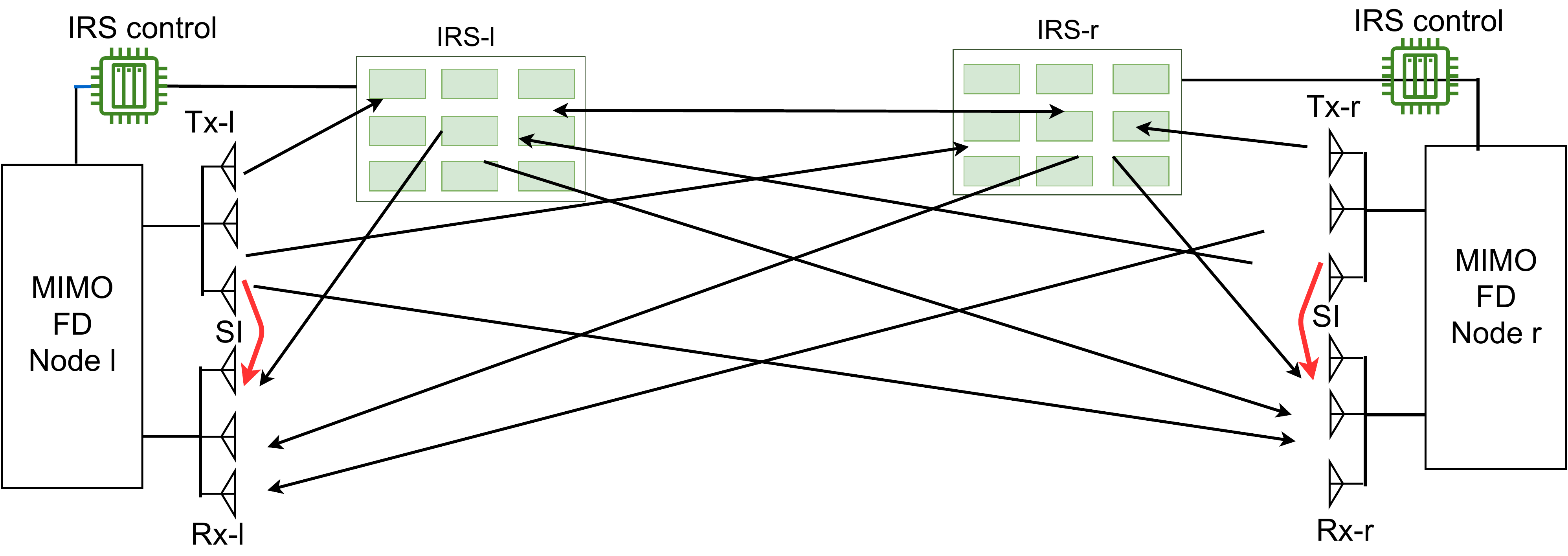}
    \caption{IRS assisted mmWave FD system with no analog beamforming.}
    \label{fig_1} \vspace{-5mm}
\end{figure}

 
We remark that the literature on FD systems with IRSs is limited only to sub-$6$ GHz. One crucial point which makes the direct application of the IRS to mmWave FD with HYBF questionable is: the analog stage of the hybrid transceivers suffers from high power consumption for phase shifters configuration and high insertion loss \cite{gao2016energy}. Our work offers a groundbreaking contribution to the field of mmWave FD systems. We consider a point-to-point MIMO scenario and propose 
to design the mmWave FD transceivers without analog beamforming and with only a very limited number of antennas. 
This approach leads to a significant reduction in the hardware cost, size and power consumption of the FD transceivers. To compensate for the massive antenna gain to overcome the propagation challenges of the mmWave, we assist each FD node with near-field (NF-) IRS, which controls the UL and DL channels and assists with strong passive SIC as shown in Fig. \ref{fig_1}. A novel joint active and passive beamforming design to maximize the weighted sum rate (WSR) is presented for the NF-IRSs-assisted FD systems. Results show that mmWave FD systems equipped with a few antennas and assisted with NF-IRSs can significantly outperform the massive MIMO (mMIMO) mmWave FD systems operating with hundreds of antennas, which makes the proposed novel transceiver design highly desirable. 



 
\vspace{-1mm}
\section{System Model} \label{sistema}
\vspace{-1mm}
We consider a mmWave point-to-point FD communication system consisting of two MIMO FD nodes, indicated with indices $l$ and $r$. They are assisted with one NF-IRS each, denoted with the indices $i_l$, and $i_r$, respectively. Both IRSs are assumed to be placed in such a way that the link between the SI channel and the other node is in line-of-sight (LoS). Each IRS operates in the NF for the assisted FD node and in the far field (FF) for the other node. The MIMO FD node $l$ is assumed to be equipped with $M_l$ transmit and $N_l$ receive antennas, while node $r$ is assumed to be equipped with $M_r$ transmit and $N_r$ receive antennas. Let $\bmV_{l} \in \mathbb{C}^{M_l \times d_l}$ and  $\bmV_{r} \in \mathbb{C}^{ M_r \times d_r}$ denote the digital beamformers of the two nodes, for the unitary-variance data streams $\bms_l \in \mathbb{C}^{d_l \times 1}$ and $\bms_r \in \mathbb{C}^{d_r \times 1}$, respectively. The IRS $i_l$ has $L_r \times  L_c$ elements, and IRS $i_r$ has $R_r \times R_c$, elements. Let $\bm{\phi}_l$ and $\bm{\phi}_r$ denote vectors collecting  the phase shift response of the IRSs $i_l$ and $i_r$, respectively. The elements of $\bm{\phi}_l$ and $\bm{\phi}_r$ at position $i$ and $j$ are phasors of the form $\bm{\phi}_l(i) = e^{i\theta_i^l}$ and $\bm{\phi}_r(j) =e^{i\theta_j^r}$, respectively. Let $\mathbf{\Phi}_l = \mbox{diag}(\bm{\phi}_l)  \in \mathbb{C}^{L_r L_c \times L_r L_c}$ and $\mathbf{\Phi}_r = \mbox{diag}(\bm{\phi}_r)  \in \mathbb{C}^{R_r R_c \times R_r R_c}$ denote the diagonal matrices containing the phase responses of the IRSs $i_l$ and $i_r$, respectively.
 
We assume perfect channel state information, which can be obtained by applying techniques like \cite{wang2020channel}. Let $\bmH_{i_l,l} \in \mathbb{C}^{L_r L_c \times M_l}$ and $\bmH_{i_r,l} \in \mathbb{C}^{R_r R_c \times M_l}$ denote the direct channels from the FD node $l$ to its NF IRS $i_l$ and to the other IRS $i_r$, operating in FF, respectively. A similar notation holds for node $i_r$.  
The channel matrices from IRSs $i_l$ and $i_r$ to the receive antenna array of FD node $l$ are  $\bmH_{l,i_l} \in \mathbb{C}^{N_l \times L_r L_c} $ and $\bmH_{l,i_r} \in \mathbb{C}^{N_l \times R_r R_c} $, respectively.
The channel matrices from the IRSs $i_l$ and $i_r$ to the receive antenna array of node $r$ are  $\bmH_{r,i_l} \in \mathbb{C}^{N_r \times L_r L_c} $ and $\bmH_{r,i_r} \in \mathbb{C}^{N_r \times R_r R_c} $, respectively. Let $\bmH_{i_r,i_l} \in \mathbb{C}^{R_r R_l \times L_r L_c} $ denote the channel from IRS $i_l$ to the IRS $i_r$, while the channel from IRS $i_r$ to IRS $i_l$ is denoted with $\bmH_{i_r,i_l} \in \mathbb{C}^{R_r R_c \times L_r L_c}$. Due to reciprocity, the channel from IRS $r$ to IRS $l$ is given by $\bmH_{i_l,i_r}=\bmH_{i_r,i_l}^T$. The direct channels from node $l$ to node $r$ and from node $r$ to node $l$ are denoted with $\bmH_{r,l} \in \mathbb{C}^{N_r \in M_l}$ and $\bmH_{l,r} \in \mathbb{C}^{N_l \in M_r}$, respectively. The SI channel for the MIMO nodes $l$ and $r$ are denoted with $\bmH_{l,l} \in \mathbb{C}^{N_l \in M_l}$ and $\bmH_{r,r} \in \mathbb{C}^{N_r \in M_r}$, respectively.
Let $\bma_l(\phi_{l}^{n_p,n_c})$ and $\bma_{r}^T(\theta_r^{n_p,n_c})$ denote the receive and transmit antenna array response at the MIMO node $l$ and $r$, respectively, with angle of arrival (AoA) $\phi_{l}^{n_p,n_c}$ and angle of departure (AoD) $\theta_r^{n_p,n_c}$. The direct channel $\bmH_{l,r}$ from node $r$ to node $l$ in mmWave can be modeled as \cite{sheemar2022practical}  \vspace{-1mm}
\begin{equation} 
    \bmH_{l,r} = \sqrt{\beta_{l,r}} \sqrt{\frac{M_r N_l}{N_c N_{p}}} \sum_{n_c = 1}^{N_c} \sum_{n_p = 1}^{N_p} \alpha_{l,r}^{(n_p,n_c)} \bma_l(\phi_{l}^{n_p,n_c}) \bma_{r}^T(\theta_r^{n_p,n_c}), \label{channel_model}
\end{equation} 
where $N_c$ and $N_{p}$ are the number of clusters and  rays, respectively, and $\alpha_{l,r}^{(n_p,n_c)} \sim \mathcal{CN}(0,1)$. Thus, the channels have amplitude and phase distributed according to the Rayleigh and uniform distribution, respectively. Note that, compared to \cite{sheemar2022practical}, \eqref{channel_model} is scaled by $\beta_{l,r} = D_{l,r}/D_{l,r}$, which captures its dependency on the distance, with $D_{l,r}$ being a distance between the centers of the MIMO FD nodes $l$ and $r$: this is equal to one for the direct link \eqref{channel_model} and will be larger than 1 for all the channels closer than  $D_{l,r}$, emphasizing the stronger channel gain. Namely, for any channel matrix $\bmH_{m,n}$, its distance-dependent scale factor is $\beta_{m,n} = D_{l,r}/D_{m,n}$, with $D_{m,n} \leq D_{l,r}, \forall m,n.$ 
The remaining FF channels $\bmH_{r,l}$, $\bmH_{i_r,l}$, $\bmH_{l,i_r}$, $\bmH_{i_r,l}$, $\bmH_{i_l,r}$, $\bmH_{i_l,i_r}$, and $\bmH_{i_r,i_l}$ can be modeled similarly to \eqref{channel_model}. The SI channel for node $l$ can be modeled as \cite{sheemar2022practical}
\begin{equation} \label{SI_Channel}
    \mathbf{H}_{l,l} = \sqrt{\beta_{l,l}} \Big[ \sqrt{\frac{\kappa_l}{\kappa_l+1}} \mathbf{H}_l^{LoS} + \sqrt{\frac{1}{\kappa_l+1}} \mathbf{H}_l^{ref} \Big],
\end{equation}
where  scalar $\kappa_l$ is the Rician factor,
$\mathbf{H}_l^{LoS}$ and $\mathbf{H}_l^{ref}$ denote 
the LoS and reflected contributions for the SI channel, respectively, and  scalar $\beta_{l,l} = D_{l,r}/D_{l,l}$ is the distance-dependent average channel gain, with $D_{l,l}$ the distance between the centers of the transmit and receive array of the MIMO FD node $l$. The channel response $\mathbf{H}_l^{ref}$ can be modeled as in \eqref{channel_model}. For the line-of-sight (LoS) path $\mathbf{H}_l^{LoS}$, we consider a spherical wave-front, which allows us to model it as \cite{sheemar2022practical}
\begin{equation} \label{SI_LOS_model}
    \bmH_l^{LoS}(m,n) = \frac{\rho}{r_{m,n}} e^{-i 2 \pi \frac{d_{m,n}}{\lambda}},
\end{equation}
where $\lambda$ denotes the wavelength and the power normalization constant $\rho$ assures that  $\mathbb{E}(||\bmH_l^{LoS}||_F^2)=M_l N_l$; the scalar $d_{m,n}$ denotes the distance between $m$-th receive and $n$-th transmit antenna of node $l \in \mathcal{F}$. Also, the channels $\bmH_{i_l,l}$, $\bmH_{l,i_l}$, $\bmH_{i_r,r}$, and $\bmH_{r,i_r}$, which involve NF-IRS, can be modelled as \eqref{SI_LOS_model}.

\subsection{Problem Formulation}
Let $\Tilde{\bmH}_{l,r}, \Tilde{\bmH}_{r,l}, \Tilde{\bmH}_{l,l}$ and $\Tilde{\bmH}_{r,r}$ denote the effective channels $\bmH_{l,r}, \bmH_{r,l}, \bmH_{l,l}$ and $\bmH_{r,r}$, respectively, including the smartly controlled paths by the IRS response, given as  
\begin{subequations} \label{eff_channels}
\begin{equation}
  \Tilde{\bmH}_{l,r} = \bmH_{l,r} + \bmH_{l,i_r} \mathbf{\Phi}_r \bmH_{i_r,r} + \bmH_{l,i_l} \mathbf{\Phi}_l \bmH_{i_l,r}  + \bmH_{l,i_l}\mathbf{\Phi}_l \bmH_{i_l,i_r} \mathbf{\Phi}_r \bmH_{i_r,r}
\end{equation}
\begin{equation}
  \Tilde{\bmH}_{r,l} = \bmH_{r,l}  + \bmH_{r,i_l} \mathbf{\Phi}_l \bmH_{i_l,l} + \bmH_{r,i_r} \mathbf{\Phi}_r \bmH_{i_r,l} + \bmH_{r,i_r} \mathbf{\Phi}_r \bmH_{i_r,i_l}  \mathbf{\Phi}_l \bmH_{i_l,l}
\end{equation}
\begin{equation}
  \Tilde{\bmH}_{l,l} = \bmH_{l,l}  + \bmH_{l,i_l} \mathbf{\Phi}_l \bmH_{i_l,l} + \bmH_{l,i_r}\mathbf{\Phi}_r \bmH_{i_r,l}   + \bmH_{l,i_r} \mathbf{\Phi}_r \bmH_{i_r,i_l} \mathbf{\Phi}_l \bmH_{i_l,l}  
\end{equation}
\begin{equation}
  \Tilde{\bmH}_{r,r} = \bmH_{r,r} + \bmH_{r,i_r} \mathbf{\Phi}_r \bmH_{i_r,r} + \bmH_{r,i_l} \mathbf{\Phi}_l \bmH_{i_l,r} + \bmH_{r,i_l} \mathbf{\Phi}_l \bmH_{i_l,i_r}  \mathbf{\Phi}_r \bmH_{i_r,r}.
\end{equation}
\end{subequations} 

Let $\bmy_l$ and $\bmy_r$ denote the total received signal at the FD nodes $l$ and $r$, respectively, which can be written as
\begin{subequations} \label{signal_model}
\begin{equation}
\begin{aligned}
        \bmy_l =\Tilde{\bmH}_{l,r} \bmV_r \bms_r +  \Tilde{\bmH}_{l,l} \bmV_l \bms_l +  \bmn_l\;\;
        \bmy_r = \Tilde{\bmH}_{r,l} \bmV_l \bms_l  +  \Tilde{\bmH}_{r,r} \bmV_r \bms_r  + \bmn_r
\end{aligned}
 \end{equation}
\end{subequations}
with $\bmn_l \sim \mathcal{CN}(\bm0, \bmI)$ and $\bmn_r \sim \mathcal{CN}(\bm0,\bmI)$ denoting the additive white Gaussian noise (AWGN) vectors with unit variance. Let $\overline{k}$ denote the indices in the set $\mathcal{F}$ except the element $k$. Let ($\bmR_{l}$) $\bmR_{\overline{l}}$ and ($\bmR_{r}$) $\bmR_{\overline{r}}$ denote the (signal and) interference plus noise covariance matrices. Let $\bmS_i = \Tilde{\bmH}_{j,i} \bmV_i  \bmV_i^H \Tilde{\bmH}_{j,i}^H,$ with $i \neq j, i \in \mathcal{F}$, denote the useful signal for node $i$. Then, we can write the covariance matrices $\bmR_{l}$ and $\bmR_{r}$ as

 \begin{subequations}
    \begin{equation}
  \bmR_{l}= \bmS_l + \Tilde{\bmH}_{l,l} \bmV_l \bmV_l^H \Tilde{\bmH}_{l,l}^H +  \bmI, 
\end{equation} 
\begin{equation}
     \bmR_{r}= \bmS_r + \Tilde{\bmH}_{r,r} \bmV_r \bmV_r^H \Tilde{\bmH}_{r,r}^H + \bmI,
\end{equation}
 \end{subequations}

 and $\bmR_{\overline{l}} = \bmR_{l} - \bmS_l$ and $\bmR_{\overline{r}} = \bmR_{r} - \bmS_r$. The WSR maximization problem for the NF-IRSs-assisted mmWave MIMO FD system, under the sum-power constraint for the FD nodes and the unit-modulus constraint for the NF-IRSs can be formally stated as
\begin{subequations} \label{org_cst}
\begin{equation} \vspace{-3mm}   \underset{\substack{\bmV_l,\bmV_r,\\\mathbf{\Phi}_l,\mathbf{\Phi}_r}}{\text{max}} \quad w_l \mbox{ln}\big[\mbox{det} \big(\bmR_{\overline{l}}^{-1} \bmR_{l} \big)\big] + w_r \mbox{ln} \big[\mbox{det}\big(\bmR_{\overline{r}}^{-1} \bmR_{r} \big)\big]
    \end{equation} 
\begin{equation} \label{c1_WSR} \vspace{-1mm}
\text{s.t.} \quad  \mbox{Tr}\big(\bmV_{k} \bmV_{k}^H \big) \leq  p_k, \quad \forall k \in \mathcal{F},
\end{equation}
\begin{equation} \label{c2_WSR} \vspace{-1mm}
  \quad  \quad  |\phi_{r}(k)| = 1 \; \& \; |\phi_{m}(j)| = 1, \forall k,m,
\end{equation}
 \end{subequations}
where $w_l$ and $w_r$ denote the rate weights for node $l$ and $r$, respectively, $p_l$ and $p_r$ denote the total sum power constraint.

 \section{Joint Active and Passive Beamforming} \label{BF}

The WSR maximization problem stated above is not concave, and finding its global optimum is challenging. To obtain a simpler solution, we adopt the weighted minimum mean squared error (WMMSE) method \cite{sheemar2022hybrid}. 

Let $\bmF_l$ and $\bmF_r$ denote the combiners used at the nodes $l$ and $r$, respectively, such that the estimated data streams can be written as 
     $\hat{\bms}_r  =  \bmF_l \bmy_r$ and $ \hat{\bms}_l  =  \bmF_r \bmy_l, $ , respectively.
Let $\bmE_l$ and $\bmE_r$ denote the mean squared error (MSE) covariance matrices for nodes $l$ and $r$, respectively, given as  
\begin{subequations} \label{errore_cov}
\begin{equation} \label{err_l}
       \bmE_l = \mathbb{E}[(\bmF_l \bmy_l - \bms_r)(\bmF_l \bmy_l - \bms_r)^H], 
\end{equation}
\begin{equation}
        \bmE_r = \mathbb{E}[(\bmF_r \bmy_r - \bms_l)(\bmF_r \bmy_r - \bms_l)^H].
\end{equation}
\end{subequations} 
We assume that  combiners $\bmF_l$ and $\bmF_r$ are optimized based on the minimum MSE (MMSE) criteria, 
which leads to the following expressions for the optimal MMSE combiners
\begin{subequations} \label{comb}
\begin{equation} 
    \bmF_l = \bmV_r^H \Tilde{\bmH}_{l,r}^H (\Tilde{\bmH}_{l,r} \bmV_r \bmV_r^H \Tilde{\bmH}_{l,r}^H + \Tilde{\bmH}_{l,l} \bmV_l \bmV_l^H \Tilde{\bmH}_{l,l}^H +  \bmI)^{-1},
\end{equation}
\begin{equation}
    \bmF_r = \bmV_l^H \Tilde{\bmH}_{r,l}^H (\Tilde{\bmH}_{r,l} \bmV_l \bmV_l^H \Tilde{\bmH}_{r,l}^H + \Tilde{\bmH}_{r,r} \bmV_r \bmV_r^H \Tilde{\bmH}_{r,r}^H +   \bmI)^{-1}.
\end{equation}
\end{subequations}

By plugging \eqref{comb} in \eqref{errore_cov}, we get that
\begin{equation} \label{opt_comb_err_l}
    \bmE_l = (\bmI + \bmV_r^H \Tilde{\bmH}_{l,r}^H\bmR_{\overline{l}}\Tilde{\bmH}_{l,r}\bmV_r)^{-1}, \;\;
    \bmE_r = (\bmI + \bmV_l^H \Tilde{\bmH}_{r,l}^H\bmR_{\overline{r}}\Tilde{\bmH}_{r,l}\bmV_l)^{-1}.
\end{equation}
%
%
Now, maximizing the WSR is equivalent to minimizing the weighted MSE error as
\begin{subequations}
\begin{equation} \vspace{-3mm}
\underset{\substack{\bmV_l,\bmV_r,\\\mathbf{\Phi}_l,\mathbf{\Phi}_r}}{\min} \quad \mbox{Tr}(\bmW_l \bmE_l) + \mbox{Tr}(\bmW_r \bmE_r)
    \end{equation}\vspace{-3mm}
\begin{equation} 
\text{s.t.} \eqref{c1_WSR}-\eqref{c2_WSR}
\end{equation} 
\end{subequations} \vspace{-1mm}
where $\bmW_i$ and $\bmW_l$ are the constant weight matrices associated with the MIMO FD nodes $l$ and $r$, respectively, which can be computed as \cite{sheemar2022hybrid}
\begin{equation} \label{rate_weights}
     \bmW_l =\frac{w_l}{\mbox{ln}\;2} \bmE_l^{-1}, \quad \bmW_r = \frac{w_r}{\mbox{ln}\;2}  \bmE_r^{-1}.
\end{equation}

\subsection{Active Digital Beamforming}
We first tackle the digital beamforming optimization problem by assuming the remaining variables to be fixed as

\begin{subequations} \label{dtre}
\begin{equation}
    \underset{\substack{\bmV_l,\bmV_r}}{\min} \quad \mbox{Tr}(\bmW_l \bmE_l) + \mbox{Tr}(\bmW_r \bmE_r),
    \end{equation} \vspace{-2mm}
\begin{equation}
\text{s.t.}  \quad    \mbox{Tr}\big(\bmV_{k} \bmV_{k}^H \big) \leq  p_k, \quad \forall k \in \mathcal{F}.
\end{equation}
\end{subequations}
By taking the partial derivative of the Lagrangian function of \eqref{dtre} with respect to the digital beamformers $\bmV_i$, where $i \in \mathcal{F}$ can be written as
 
 \vspace{-2mm}
\begin{equation} \label{BFs}
\begin{aligned}
    \bmV_i = &\big( \bmX_i + \mu_i \bmI \big)^{-1}  \Tilde{\bmH}_{j,i}^H\bmF_{j}^H \bmW_j,
\end{aligned}
\end{equation}
where $i \neq j$ and $i,j \in \mathcal{F}$, and the matrix 
$\bmX_i$ is defined as
\begin{equation}
    \bmX_i = \Tilde{\bmH}_{j,i}^H  \bmF_j^H \bmW_j \bmF_j \bmH_{j,i}+ \bmH_{i,i}^H \bmF_i^H \bmW_j \bmF_i \bmH_{i,i},
\end{equation}
and scalars $\mu_i$ denote the Lagrange multiplier for the sum-power constraint of the MIMO FD nodes $i \in \mathcal{F}$. To find the Lagrange multiplier, we can  consider the singular value decomposition of $\bmX_i =  \bmU_i \mathbf{\Sigma}_i \bmU_i^H,$ where $i \in \mathcal{F}$, and write the power constraint as
 \vspace{-3mm}
\begin{equation}
    \mbox{Tr}(\bmV_i \bmV_i^H) = \sum_{k=i}^{N_l} \frac{\bmG_i(k,k)}{(\mu_i + \mathbf{\Sigma}_i(k,k))^2} = p_i,
\end{equation}
where the matrices $\bmG_i = \bmU_i^H \Tilde{\bmH}_{j,i}^H \bmF_{j}^H \bmW_j \bmW_j \bmF_{j} \Tilde{\bmH}_{j,i} \bmU_i$,
with $i \neq j,$ and $i,j\in\mathcal{F}$. To find the optimal values of $\mu_i$, $\forall i$, we adopt the Bisection method.

\subsection{Passive Beamforming With IRSs}
We now consider the optimization of $\mathbf{\phi}_i$, $\forall i$, for which the MMSE optimization problem can be written as 
  \begin{subequations}
\begin{equation}  \label{IRS_i_opt_restatedoo}
    \underset{\substack{\mathbf{\phi}_i}}{\min} \quad \bm{\phi}_i^H  \mathbf{\Sigma}_i \bm{\phi}_i +{\bms_i}^H \bm{\phi}_i^* + {\bms_i}^T \bm{\phi}_i, 
    \end{equation} 
    \vspace{-3mm}
\begin{equation} \label{c1oo}
\text{s.t.} \quad 
    |\phi_{i}(k)| = 1,  \quad \forall k.
\end{equation}
\end{subequations}
where $\mathbf{\Sigma}_i = \bmZ_i \odot \bmT_i^T$ and  $\bmZ_i$ and $\bmT_i^T$ are given as
\begin{equation}
\begin{aligned}
    \bmZ_{i} = & \bmH_{i,i_i}^H \bmA_{1,i}^H \bmW_i \bmA_{1,i} \bmH_{i,i_i} + \bmC_{2,i}^H \bmA_{1,i}^H \bmW_i \bmA_{1,i} \bmC_{2,i} \\& + \bmA_{1,i}^H \bmW_j \bmA_{1,i} \bmB_{2,i}  
    + \bmH_{j,i_i}^H \bmA_{1,j}^H \bmW_j \bmH_{j,i_i},
\end{aligned}
\end{equation}
\begin{equation}
\begin{aligned}
     \bmT_i = & \bmA_{2,i} \bmV_j \bmV_j^H \bmA_{2,i}^H + \bmH_{i_i,i} \bmV_i \bmV_i^H \bmH_{i_i,i}^H + \bmH_{i_i,i} \bmV_i \bmV_i^H \bmH_{i_i,i}^H  \\&+ \bmD_{2,i} \bmV_j \bmV_j^H \bmD_{2,i}^H.
\end{aligned}
\end{equation}
   
 in which the auxiliary matrices are defined as

\begin{subequations}
\begin{equation}
     \bmA_{1,i} = \bmH_{i,j} + \bmH_{i,i_j} \mathbf{\Phi}_j \bmH_{i_j,j}, \quad \bmA_{2,i} = \bmH_{i_i,j} + \bmH_{i_i,i_j} \mathbf{\Phi}_j \bmH_{i_j,j}
 \end{equation}
 \begin{equation}
     \bmB_{1,i} = \bmH_{j,i} + \bmH_{j,i_j} \mathbf{\Phi}_j \bmH_{i_j,i}, \quad \bmB_{2,i} = \bmH_{j,i_i} + \bmH_{j,i_j} \mathbf{\Phi}_j \bmH_{i_j,i_i},
 \end{equation}
 \begin{equation}
     \bmC_{1,i} = \bmH_{i,i} + \bmH_{i,i_j} \mathbf{\Phi}_j \bmH_{i_j,i}, \quad \bmC_{2,i} = \bmH_{i,i_i} + \bmH_{i,i_j} \mathbf{\Phi}_j \bmH_{i_j,i_i}, 
 \end{equation}
 \begin{equation}
   \bmD_{1,i} = \bmH_{j,j} + \bmH_{j,i_j} \mathbf{\Phi}_j \bmH_{i_j,j}, \quad \bmD_{2,j} = \bmH_{i_j,i} + \bmH_{i_j,i_i} \mathbf{\Phi}_i \bmH_{i_i,i}.
 \end{equation}
\end{subequations}
and in \eqref{IRS_i_opt_restatedoo}, vector $\bms_i$ is made of the diagonal elements of the matrix 
\begin{equation}
\begin{aligned} \label{matrices_Sl}
       \bmS_{i} = & \bmA_{2,i} \bmV_j \bmV_j^H \bmA_{1,j}^H \bmF_i^H \bmW_i \bmF_i \bmH_{i,i_i} + \bmH_{i,i} \bmV_i \bmV_i^H \bmC_{1,i}^H \bmA_i^H \bmW_i \bmA_i \bmC_{2,i} \\&- \bmA_{2,i} \bmV_j \bmW_i \bmA_i \bmH_{l,i_i}  + \bmH_{i_i,i} \bmV_i \bmV_i^H \bmB_{1,i}^H \bmA_i^H \bmW_j \bmA_j \bmB_{2,i} \\& + \bmD_{2,i} \bmV_j \bmV_j^H \bmD_{1,i}^H \bmA_j^H \bmW_j \bmA_j \bmH_{j,i_i} - \bmH_{i_i,i} \bmV_i \bmW_j \bmA_j \bmB_{2,i}.
\end{aligned}
\end{equation}

Problem \eqref{IRS_i_opt_restatedoo}  is non-convex due to the unit-modulus constraint. To render a feasible solution, we adopt the majorization-maximization optimization method \cite{pan2020multicell}. Its objective is to solve a difficult problem through a series of more tractable sub-problems made of the touching upper bounds at each iteration.  For the problem of the form \eqref{IRS_i_opt_restatedoo}, a simple upper bound can be constructed as \cite{pan2020multicell}
\begin{equation} \label{UB_r} \vspace{-1mm}
     g(\bm{\phi}_i|\bm{\phi}_i^{(n)}) = 2 \mbox{Re}\{{\bms_i}^H \bmq_i^{(n)}\} + o_i,
 \end{equation}
where $o_i$ denotes constant terms, and $\bmq_i^{(n)}$ is given by 
\begin{equation} \vspace{-1mm}
    \bmq_i^{(n)} = (\lambda_i^{max} \bmI - \mathbf{\Sigma}_i) \bm{\phi_i}^{(n)} - {\bms_i}^*,
\end{equation}
and $\lambda_i^{max}$ denotes the maximum eigenvalues of $\mathbf{\Sigma}_i$. Given the upper bound, problem \eqref{IRS_i_opt_restatedoo} can be restated as  
 \begin{subequations}  \label{IRS_r_optimized_ert}
\begin{equation}
    \underset{\substack{\bm{\phi}_i}}{\min} \quad  2 \mbox{Re}\{{\bms_i}^H \bmq_i^{(n)}\},
    \end{equation}  
    \vspace{-4mm}
\begin{equation} \label{c1_trew}
\text{s.t.} \quad 
   |\bm{\phi}_{i}(k)| = 1,  \quad \forall k.
   \end{equation}
 \end{subequations}
For \eqref{c1_trew}, the optimal solution is given as
\begin{equation} \label{solution_phi_r}
    \bm{\phi_i}^{(n+1)} = e^{i \angle\bmq_i^{(n)}}.
\end{equation}
 
At each iteration, when the digital beamformers are fixed, the optimization of $\mathbf{\Phi}_i, \forall i \in \mathcal{F}$ consists in updating their response iteratively until convergence by solving a series of more tractable optimization problems, and the formal procedure is provided in Algorithm~$1$. The joint active and passive beamforming procedure for the mmWave FD system is formally stated in Algorithm $2$.

\begin{algorithm}[t]  
\caption{Optimization of IRS $i \in \mathcal{I}$ }\label{alg_1}
\textbf{Initialize:}  iteration index $n=1$, accuracy $\epsilon$.\\
\textbf{Evaluate:} $f(\bm{\phi}_i(0))$.\\
\textbf{Repeat until convergence}
\begin{algorithmic}
\STATE \hspace{0.001cm} Calculate $\bmq_i^{(n)} = (\lambda_i^{max} \bmI - \mathbf{\Sigma}_i) \bm{\phi}_i^{(n)} - {\bms_i}^*$
\STATE  \hspace{0.001cm} Update $\bm{\phi}_i^{(n+1)}$ with  $\bm{\phi}_i^{(n+1)}=\bm{\phi_r}^{(n+1)} = e^{i \angle\bmq_i^{(n)}}$.\\
\STATE  \hspace{0.001cm} \textbf{if} $|f(\bm{\phi}_i^{(n+1)}) - f(\bm{\phi}_i^{(n)})|/f(\bm{\phi}_i^{(n+1)}) \leq \epsilon$
\STATE  \hspace{0.4cm} Stop and return $\bm{\phi}_i^{(n+1)}$.
\STATE  \hspace{0.001cm} \textbf{else} n=n+1 and repeat.
\end{algorithmic} \label{Alg_1}
\end{algorithm}

\begin{algorithm}[t]  
\caption{WSR maximization for MIMO FD with NF-IRSs}\label{alg_3}
\textbf{Initialize:}  accuracy $\epsilon$, beamformers and combiners.\\
\textbf{Repeat until convergence}
\begin{algorithmic}
\STATE for $\forall i \in \mathcal{F}$\\
\STATE \hspace{0.3cm} Update $\bmF_i$ with \eqref{comb}.\\
\STATE \hspace{0.3cm} Update $\bmW_i$ with \eqref{rate_weights}.\\
\STATE \hspace{0.3cm} Update $\bmV_i$ with \eqref{BFs}.\\
\STATE \hspace{0.3cm} Update $\mathbf{\Phi}_i$ with Algorithm  \ref{Alg_1}.\\
\STATE  \textbf{if} convergence condition is satisfied\\
\STATE  \hspace{0.4cm} Stop and return the optimized variables.
\STATE  \textbf{else} repeat.
\end{algorithmic} 
\end{algorithm}  
 
\subsection{Convergence}
 The proof for Algorithm \ref{alg_3} can be derived by combining the proof for WMMSE for digital beamformers and majorization-minimization technique for IRS \cite{pan2020multicell}. Due to space limitations, we consider omitting the proof.

\section{Numerical Results} \label{risultati}
In this section, we present simulation results to evaluate the performance of the proposed joint active and passive beamforming for mmWave FD systems. Note that digital beamforming designs represent the upper bounds for the HYBF designs. Therefore, we compare our solution against the mMIMO fully digital systems, which provide the maximum achievable gain for the HYBF designs. For comparison, we define the following benchmark schemes:
1) A mMIMO HD system with fully digital beamforming capability with $100$ transmit and $50$ receive antennas for each MIMO HD node, which is denoted as HD-$100\times 50$, 2) A mMIMO FD system with fully digital beamforming capability with $100$ transmit and $50$ receive antennas for each MIMO FD node, denoted as FD-$100\times 50$.

We define the signal-to-noise ratio (SNR) for the NF-IRSs assisted FD system as SNR = $p_r/\sigma_r^2 = p_l/\sigma_l^2$, where $p_i$ and $\sigma_i^2$ denote the total transmit power and noise variance for node $i$, respectively, with $i=l$ or $i=r$. We assume that the considered systems operate at the frequency of $30$~GHz, i.e., $\lambda = 10$~mm. The noise variance is set to be $\sigma_i^2=1$ and the total transmit power is chosen to meet the desired SNR. For the NF-IRS assisted FD system, the number of antennas is chosen to be $5$ times less, i.e., only $M_l=M_r=20$ transmit and $N_l=N_r=10$ receive antennas, respectively, and the number of data streams is $d_l=d_r=2$. The FD nodes are assumed to be $D_{l,r}= 200$~m far. We consider the NF-IRSs of size $10 \times 10$, $20 \times 20$, and $30 \times 30$ for both the MIMO FD nodes, with the center between two IRS elements located at $\lambda/2$. They are placed at the minimum distance of $3$~m from their FD nodes, which is varied up to $90$~m. We assume that both the FD nodes are equipped with ULAs, and their transmit and the receive arrays are assumed to be separated by a distance $D_b=20$~cm, with a relative angle $\Theta_b = 180^\circ$ and $r_{m,n}$ in \eqref{SI_LOS_model} is set given $D_b$ and $\Theta_b$ as in \cite[(9)]{sheemar2022practical}. The Rician factor is set as $\kappa_b=1$ and the rate weights are chosen as $w_{l}=w_{r}=1$. The number of paths and the number of clusters is set as $N_{c,b}=N_{p,b}=3$ and the AoA $\theta_{r}^{n_{p},n_{c}}$ and AoD $\phi_{l}^{n_{p},n_{c}}$ are assumed to be uniformly distributed in the interval $\mathcal{U}\sim [-30^\circ,30^\circ]$.  

Let $\hat{x},\hat{y}$, and $\hat{z}$ denote the $3$ versors on the three-dimensional space and any point on the $3$ dimensional vector space can be written as $(x \hat{x},y \hat{y}, z \hat{z})$. We assume the FD nodes with ULAs to be aligned with $\hat{x}$ direction and the first transmit antenna of the FD nodes $l$ and $r$ are assumed to be placed in the positions $(0,0,0)$ and $(0,D_{l,r\hat{y}},0)$, respectively, with other antennas placed in positions $x+$. Both the NF-IRSs are assumed to be at the same distance from their FD nodes, and placed on the x-y plane with the first element of the NF-IRSs $i_l$ and $i_r$ placed in positions $(0,D_{IRS}\hat{y},0)$ and $(0,D_{r,l}-D_{IRS}\hat{y},0)$, respectively. 
Note that when the IRS operate in the NF of the FD node, joint passive SIC and intelligent control of the UL and DL channels also depend on the orientation of the NF-IRSs.
Namely, the NF-IRSs can provide better SIC by jeopardizing the UL and DL channel control if they are more oriented towards the receive antennas of the NF FD node, or vice versa. The chosen orientation of the NF-IRSs in our setup provides an optimal trade-off between the passive SIC capabilities of NF-IRSs and control over the UL and DL channels.


\begin{figure}
    \centering
    \includegraphics[width=0.7\columnwidth,height=4cm]{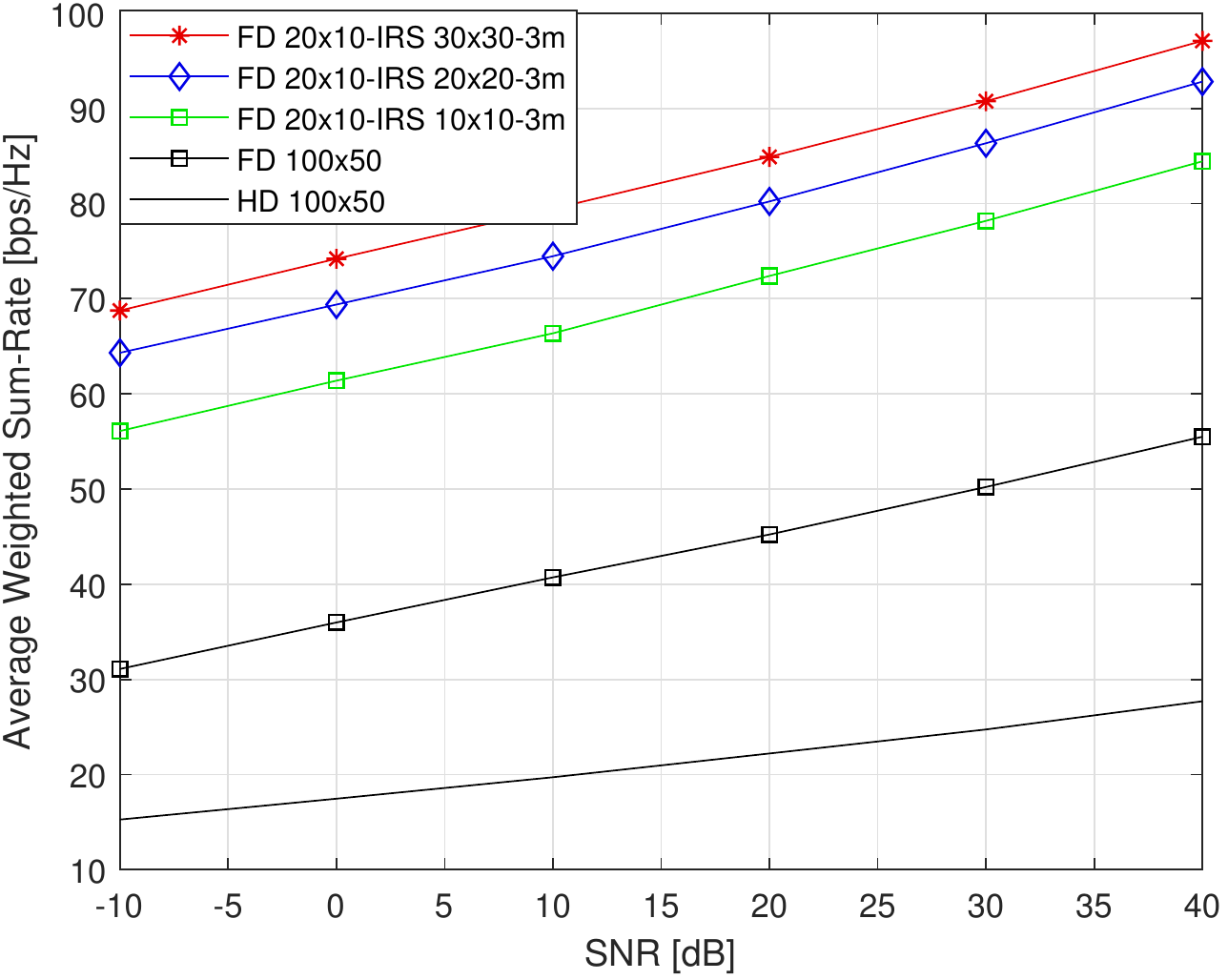}\vspace{-2.5mm}
    \caption{Average WSR as a function of SNR with NF-IRSs of size $30\times 30$, $20\times 20$, and $10\times 10$, placed at a distance of $3$~m.}
    \label{fig:_IRS_Fig1}\vspace{-3mm}
\end{figure}

\begin{figure}
    \centering
    \includegraphics[width=0.7\columnwidth,height=4cm]{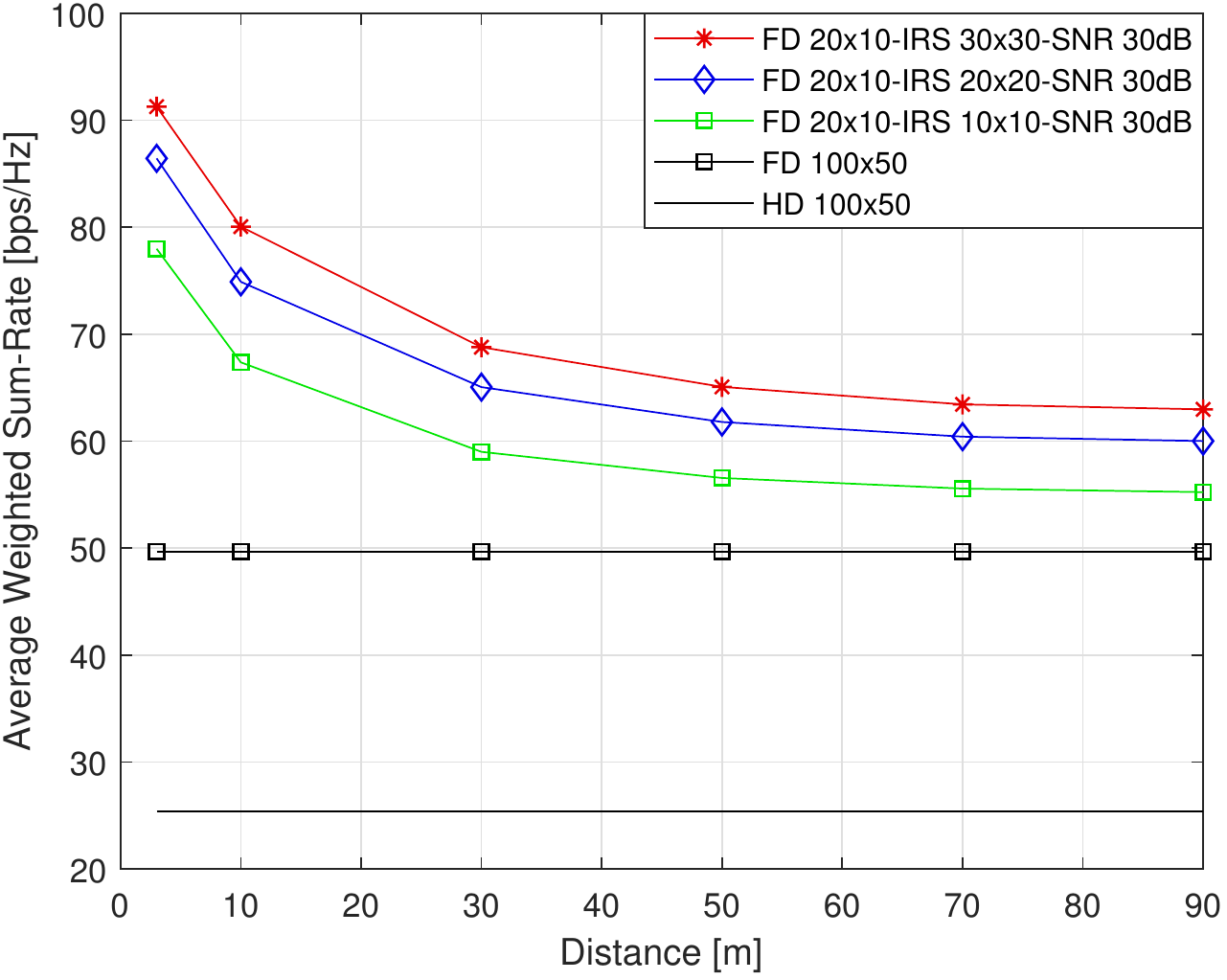}\vspace{-2.5mm}
    \caption{Average WSR as a function of the distance between the FD nodes and the NF-IRSs at SNR$=30$~dB.}
   \label{fig:_IRS_Fig4} \vspace{-3mm}
\end{figure}

Fig.~\ref{fig:_IRS_Fig1} shows the achieved average WSR as a function of the SNR for the FD systems assisted with  NF-IRSs placed at $3$~m from the transmit array.
We can see that despite using $5$ times fewer antennas than the mMIMO FD systems, the proposed approaches significantly outperform the benchmark schemes. Namely, with NF-IRSs of size $30 \times 30$, they can achieve $\sim 4$ times higher gain than the traditional mMIMO HD system. This is mainly due to the fact that NF-IRSs are able to very strongly suppress the SI due to very small path-loss and thus provide a very high effective signal-to-SI plus noise ratio.


Fig.~\ref{fig:_IRS_Fig4} shows the performance of the proposed design, as a function of the distance between the FD node and their NF-IRSs, which is varied in the interval $3-90$~m, with the two FD nodes operating at the distance of $200$~m at SNR$=30~$dB. We can see that the proposed novel transceiver design without the analog beamforming stage for the mmWave FD system fully reaps the benefits of the IRSs when they operate in the NF. Placing the IRSs far from the mmWave FD systems leads to small performance gains and therefore a combination of analog beamforming with a massive number of antennas might be required, which may result to be a costly and highly energy inefficient. 
On the other hand, with the proposed approach, no HYBF is required and the mmWave FD nodes can operate with a few active elements, thus paving the path towards more efficient next-generation mmWave FD systems.

\section{Conclusions} \label{Conc}
 This paper proposes a novel transceiver design for the mmWave FD system with a few number of antennas, without the analog beamforming part, compared to the conventional mMIMO solution with HYBF to overcome the propagation challenges of the mmWave band. Each FD node is then assisted with an NF-IRS which aims to smartly control the UL and DL channel while performing very strong passive SIC. A novel joint active and passive beamforming solution is derived. Results demonstrate that the novel mmWave FD transceiver design is very promising and can achieve significantly higher gains while being highly cost and energy efficient.

\section*{Acknowledgement}
This work has been supported by the Luxembourg National
Research Fund (FNR) project, titled Reconfigurable Intelligent Surfaces for Smart Cities (RISOTTI).

\bibliographystyle{IEEEtran}
\bibliography{main}

\end{document}